\magnification=1200
\baselineskip=20 pt

\hfill{hep-ph 0102176}

\centerline{\bf Implications of BNL measurement of $\delta a_{\mu}$
on scalar leptoquark mass and coupling}

\vskip 1 true in

\centerline{\bf Uma Mahanta}

\centerline{\bf Mehta Research Institute}

\centerline{\bf Chhatnag Road, Jhusi}

\centerline{\bf Allahabad-211019, India}

\vskip .4 true in

\centerline{\bf Abstarct}

Recently BNL  have measured the muon magnetic moment anomaly with
increased precision. 
The world average experimental value at present shows a
discrepancy of $43(16)\times 10^{-10}$ from the Standard 
Model [SM] value. In this paper we investigate the implications of this
difference on a class of scalar leptoquark interactions to SM
quark-lepton pair.
 We find that for
 leptoquarks in the few hundred Gev range the BNL muon anomaly could 
arise from leptoquark couplings that are much smaller than the
electromagnetic coupling.
 We  also find that the BNL value for the muon anomaly leads to an
unambiguous prediction for the electric dipole moment of the muon and
a bound on 
the flavor changing leptoquark coupling relevant for the
rare decay $\mu\rightarrow e\gamma $.

\vfill\eject
The BNL collaboration has reported a new improved measurement of the
muon magnetic moment anomaly [1]. The present average value shows
a discrepancy of 43 (16) $\times 10^{-10}$ from the estimated Standard
Model [SM] value. A variety of new physics scenarios like
extra gauge bosons, compositeness, extra fermions and supersymmetry
have been considered to explain this discrepancy [2]. The muon anomaly
had also been used in the past to constrain physics beyond the SM.
For earlier works related to this subject see Ref[3].

 In this paper
we  shall analyze the possibility that the BNL muon anomaly could be due to
elementary leptoquarks.
Leptoquarks are scalar or vector particles in 3 or 3$^*$ representation
of $SU(3)_c$ that couple to quark lepton pair.
  They can be in the triplet, doublet or singlet representation
of $SU(2)_l$. They can also carry weak hypercharge. Leptoquarks occur
in many scenarios beyond the SM e.g technicolor models, substructure
models of quarks and leptons and string inspired grand unified models like
E(6).
 Consider a scalar leptoquark which has the following  
Yukawa interactions [4] to SM fermions

$$L^{np}=(g_{Lij}{\bar q}^{ci}_R i\tau_2 l_L^j +
g_{Rij}{\bar u}^{ci}_L e^j_R )S_1
+h.c. \eqno(1)$$

Here $g_{Lij}$ and $g_{Rij}$ are the leptoquark couplings to LH
and RH leptons. i and j refer to quark and lepton generations.
$\psi ^c =C\bar {\psi}^T $ is the charge conjugated spinor.
 $S_1$ is second generation
leptoquark in the 3$^*$ representation of color. It is a  weak isoscalar
and has charge ${1\over 3}$.
The above interaction Lagrangian will generate a one loop correction to the 
magnetic moment [5] of muon which can be estimated from the following
effective Lagrangian

$$L_{eff}= \sum _i ie  m_{ui} I^i(0, 0)Re(g^*_{Li2}g_{Ri2})
 {\bar\mu} \sigma_{\mu\nu}\mu F^{\mu\nu}.\eqno(2)$$

where $Re(g^*_{Li2}g_{Ri2})$  means real part of $( g^*_{Li2}g_{Ri2})$,
 $m_{ui}$ is the mass of the $Q={2\over 3}$ quark of  ith generation and
$$I^i(0, 0)=\int{d^4l\over (2\pi )^4}{1\over (l^2-m_{ui}^2) [l^2-m_{ui}^2]
[l^2-m^2_{s}]}.\eqno(2)$$

$m_s$ is the mass of the leptoquark $S_1$.
From the above effective Lagrangian it follows that the one loop correction
to the magnetic moment of the muon is given by

$$\delta \mu=\sum_{i=1}^3 2ie m_{ui} I^i(0,0)Re(g^*_{Li2}g_{Ri2}).
\eqno(4)$$

On evaluating the the loop integral we find that
 $I^i(0,0)=-{i\over 16\pi^2 m^2_s}[-{1\over 1-r_i}-{\ln r_i\over 
(1-r_i)^2}]$ where  $r_i={m^2_{ui}\over m^2_s}$.

The contribution of the up quark to $\delta \mu$ can be neglected firstly 
because its mass is small and secondly it involves off-diagonal couplings
between first and second generation. The contribution of the top quark
however can be comparable to that of the charm quark because its large mass
can partly
compensate for the small off diagonal couplings between the second and
third generation fermions.

$$ a_{\mu}^{np}\approx \sum_{i=2}^3{ m_{ui} m_{\mu}\over 4\pi^2 m^2_{s}}
[-{1\over 1-r_i}-{\ln r_i\over (1-r_i)^2}]  Re(g^*_{Li2}
g_{Ri2}).\eqno(5)$$

$S_1$ being a second leptoquark its couplings $g_{L22}$ and $g_{R22}$
 to second generation
fermions are expected to be strongest. Whereas
 the flavor violating couplings
$g_{L32}$ and $g_{R32}$ are expected to be hierarchically smaller.
In the numerical estimates presented here we shall assume that
$Re(g^*_{L32}g_{R32})\approx .01 Re(g^*_{L22}g_{R22})$ i.e. a Cabibbo
type suppression.
 We find that
for these values of the flavor offdiagonal couplings the top contribution
is much smaller than the charm contribution. Since the BNL value of the
muon anomaly allows one to find bounds on the leptoquark mass
for given values of the couplings but not both,
we shall introduce an effective scale $\Lambda$ for expressing leptoquark
contribution to $a_{\mu}^{np}$ where

$${1\over \Lambda^2}={ Re (g^*_{L22}g_{R22})\over m_s^2} 
[-{1\over (1- r_2)}-{ln r_2
\over (1-r_2)^2}].\eqno(6)$$

The introduction of $\Lambda$ can be looked upon as a convenient
normalization of the leptoquark contribution to $a_{\mu}^{np}$.
Equating  $a_{\mu}^{np}$ to the central value for $\delta a_{\mu}$
we get $\Lambda \approx $ 890 Gev. The new value of $\delta a_{\mu}$
is important because of two reasons. First it represents a large 
discrepancy (2.6 $\sigma $ effect)  between the SM value and the
measured value. Second the new measurement has an error one 
third of the combined previous data. Both these factors can be taken into
by determining 95\%  CL limits on $\Lambda$. For the present
world average value of $\delta a_{\mu}$ these bounds are given by
: 678 Gev $\le \Lambda \le $ 1760 Gev. The ultimate goal of the experiment 
[2] is to reduce the error to $\pm 4\times 10^{-10}$ , 
about a factor of 3.5 times
 better than the new BNL result. Even the inclusion of already existing
data from the 2000 run is expected to reduce the statistical error
by a factor of 2. If the central value and other errors are 
unaffected this would improve the bounds on $\Lambda$ to
764 Gev $\le \Lambda\le$1123 Gev.
We would like to emphasize that $\Lambda$ should not be interpreted as
the leptoquark mass. The  95\% CL  bounds on the couplings for a given
value of the leptoquark mass can be obtained from eqn (6) by substituting
the bounds for $\Lambda$ given above. For example if $m_s= $ 300 Gev
the 95\% CL bounds on the couplings are $.003 \le Re (g^*_{L22}g_{R22})
\le .020 $.

The leptoquark contribution to $(g-2)_{\mu}$  involves a chirality
flip on the internal quark line. This causes the resulting expression 
to be proportional to Re$(g^*_Lg_R)$. The most stringent
 bound on $\vert g_Lg_R \vert$
for first generation leptoquarks arises from the helicity suppressed
decay mode $\pi\rightarrow e\nu_e$ and is given by $\vert g_Lg_R \vert\le
({m_{lq}\over 30 Tev})$ [4, 6].
 The analogous decay $K\rightarrow \mu\nu_e$
for second generation leptoquarks however involves flavor changing couplings.
Further since helicity suppression is not that stringent for second generation
fermions we expect the bound on Re$ (g^*_{L22}g_{R22})$ to be much weaker
than it is for first generation. Coservatively even if we assume it to be
 of the same order as for first generation it would be consistent with
the bounds $.003\le Re (g^*_{L22}g_{R22}) \le .020 $ derived in this paper
from muon anomaly.

The direct bounds on $m_{lq}$ set by Tevatron and HERA [7] assumes that
$g_{lq}\approx e$. The Tevatron bounds arise from
 pair production of leptoquarks. These bounds
depend on the color and electroweak quantum numbers of the leptoquark.
For a second generation  scalar leptoquark having the quantum numbers
of $S_1$ the Tevatron bound is $m_{lq}> 200$ Gev. 
This limit depends
upon the assumption that $B(\mu , q)=1 $. For $B(\mu ,q)=$0.5
 the bound becomes 180 Gev.
 On the other hand
HERA sets bounds on leptoquark masses from single production.
The bound on a second generation $S_1$ leptoquark is $m_{lq}> 73$ Gev.
The bounds on the leptoquark
mass derived in this paper from the muon magnetic moment anomaly
are therefore stronger than the direct bounds set by HERA and Tevatron.

We shall now consider the implications of the BNL result on the
electric dipole moment
(EDM) of muon. Since $g^*_{2L}g_{2R}$ can in general be complex the
imaginary part of  $g^*_{2L}g_{2R}$ can contribute to the EDM of muon.
In the estimates presented below we shall consider the flavor
 diagonal leptoquark
couplings only and ignore the top contribution which is expected to
be numerically smaller.
We find that the EDM of the muon is given by
$$d^{\gamma}_{\mu}\approx  2i e m_c I(0,0)Im(g^*_{L22}g_{R22})
.\eqno(6)$$

where Im means imaginary part of the relevant quantity.
If the BNL discrepancy between the measured and the SM value of the muon 
magnetic moment anomaly is due to leptoquarks then one can derive an 
order of magnitude value for the EDM of the muon. In the SM the EDM of 
leptons vanishes to three loops and is predicted to be of the order of
1.6$({m_l\over Mev})\times 10^{-40}$ e cm [8]. The SM contribution to the 
EDM of leptons is therefore
far too small to be observed in any experiments to be
performed in near future. So any observed value of the EDM of leptons
must be due to new physics. In the SM the small value of CP violation
follows from the small value of the CKM angles but not the phase.
Assuming that the phase factor $\sin \delta$ for leptoquark couplings
is of order one we have $Im(g^*_{2L}g_{2R})\approx Re(g^*_{2L}g_{2R})$.
 The 
value of $I(0,0)Re(g^*_{2L}g_{2R})$ can be estimated from the discrepancy
in $(g-2)_{\mu}$ by BNL and is given by 6.8$\times 10^{-9}$. It then 
follows from eqn (6) that the EDM of muon is of the order of
 4.1$\times 10^{-22}$e cm.
The muon anomaly reported by BNL therefore leads to an almost (except
for a choice of phase)
unambiguous
prediction for the EDM of muon.
The predicted value however 
is three orders of magnitude smaller than the present experimental
upper bound of 3.7$\times 10^{-19}$ e cm [9] on the EDM of muon. 

The BNL measurement also has important implications for the  lepton flavor 
violating decay $\mu\rightarrow e\gamma$. Leptoquarks can have tree level
flavor violating couplings to quark-lepton pairs
 and they can induce the rare decay 
$\mu\rightarrow e\gamma$ at the one loop level. It can be shown that
the relevant effective Lagrangian is given by

$$L_{eff}=\xi \bar {e} \sigma_{\mu\nu}
 \mu F^{\mu\nu} +h.c. \eqno(7)$$

where $\xi=i e m_c I(0,0) Re(g^*_{L21}g_{R22})$. It then follows that
the decay width of the transition $\mu\rightarrow e\gamma$ is given by

$$\eqalignno{\Gamma_{\mu\rightarrow e\gamma} &={e^2 m_c^2\over \pi}\vert I(0,0)
Re(g^*_{L22}g_{R22})\vert ^2 \vert {g^*_{L21}\over g^*_{L22}}\vert ^2
m_{\mu}^3 \cr
& \approx 3.9\times 10^{-18} \vert {g^*_{L21}\over g^*_{L22}}\vert ^2 
Mev .&(8)\cr}$$

In the above we have used the average value of the muon anomaly reported by
the BNL collaboration to estimate $ \vert I(0,0)
Re(g^*_{L22}g_{R22})\vert $. The present experimantal bound [10]  on the 
branching ratio for $\mu\rightarrow e\gamma$ is $1.2 \times 10^{-11}$.
It then follows that $\vert { g^*_{L21}\over g_{R22}}\vert <6.2\times 
10^{-5}$. The BNL measurement therefore allows us to determine the
scale dependent part of the transition rate unambiguously,
leaving only a mixing angle to be determined from the branching ratio
of the rare decay.
 We find that the relevant flavor violating coupling of the
second generation leptoquark must be very
 strongly suppressed relative to
its flavor diagonal coupling.

In conclusion in this paper we have analyzed the possibility that the
 BNL muon naomaly is due to a second generation
 scalar leptoquark that couple to muon of both
chiralities. We find that the muon anomaly could arise from relatively 
light leptoquarks in the few hundred Gev range and couplings that are smaller
than the electromagnetic coupling. The bounds are consistent with all other
known bounds and in particular the ones arising from helicity suppressed
decays of light mesons.
 We have also shown that the BNL muon anoamly leads to
an unambiguous prediction for the EDM of the muon and a bound on the
flavor changing leptoquark coupling relevant for the rare decay
$\mu\rightarrow e\gamma$

\centerline{\bf References}

\item{1.} H. N. Brown et al, hep-ex/0102017.

\item{2.} A. Czarnecki and W. Marciano, hep-ph/0102122; K. Lane,
hep-ph/0102131; L. Everett, G. L. Kane, S. Rigolin and L. T. Wang,
hep-ph/0102145;
 J. Feng and K. Matchev, hep-ph/0102146; E. Baltz 
and P. Gondolo, hep-ph/0102147; U. Chattopadhyay and P. Nath,
hep-ph/0102157.

\item{3.} R. Barbieri and L. Maiani, Phys. Lett. B 117, 203 (1982);
D. A. Koswoer, L. M. Krauss and N. Sakai, Phys. Lett. B 133, 305 (1983);
R. Arnowitt, A. H. Chamseddine and P. Nath, Z. Phys. C 26, 407 (1984);
C. Arzt, M. B. Einhorn and J. Wudka, Phys. Rev. D 49, 370 (1994);
J. L. Lopez, D. V. Nanopoulos and X. Wang, Phys. Rev. D 49, 366 (1994);
U. Chattopadhyay and P. Nath, Phys. Rev. D 53, 1648 (1996); T. Moroi,
Phys. Rev. D 53, 6565 (1996); M. Carena, G. F. Giudice and C. E. Wagner,
Phys. Lett. B 390, 234 (1997); P. Nath and M. Yamaguchi, Phys. Rev. D
60, 116006 (1999); U. Mahanta and S. Rakshit, Phys. Lett. B 480, 176
(2000);
 M. L. Graesser, Phys. Rev. D 61, 074019 (2000).

\item{4.} W. Buchmuller and D. Wyler, Phys. Lett. B 177, 377 (1986);
W. Buchmuller, R. Ruckl and D. Wyler, Phys. Lett. B 191, 442 (1987).

\item{5.}U. Mahanta, Phys. Rev. D 54, 3377 (1996).

\item{6.} S. Davidson, D. Bailey and B. A. Campbell, ZPHY C61, 613 (1994).

\item{7.} D. E. Groom et al, Review of Particle Physics,
 Euro. Phys. Jour. C15, 1 (2000).

\item{8.}M. J. Booth, University of Chicago Report no. EFI-93-02,
hep-ph/9301293 (unpublished).

\item{9.} D. E. Groom et al, Review of Particle Physics,
Euro. Phys. Jour. C15, 1 (2000)

\item{10.} D. E. Groom et al, Review of Particle Physics,
 Euro. Phys. Jour. C15, 1 (2000).

\end